\documentclass[prl,aps,twocolumn,showpacs]{revtex4}
\usepackage{graphicx,dcolumn,amsmath,color,longtable}

\begin{document}

\title{Chiral asymmetry of the spin-wave spectra in ultrathin magnetic films}
%due to Dzyaloshinskii-Moriya interactions} 

\author{L.\ Udvardi}
\author{L.\ Szunyogh}\email{szunyogh@phy.bme.hu}

\affiliation{Department of Theoretical Physics, Budapest University of Technology and Economics, Budafoki \'ut 8. H1111 Budapest}

\date{\today}

\begin{abstract}
%Based on theoretical considerations and first principles calculations, 
We raise the possibility
that the chiral degeneracy of the magnons in ultrathin films
can be lifted due to the presence of Dzyaloshinskii-Moriya interactions.
By using simple symmetry arguments, we discuss under which conditions such a 
chiral asymmetry occurs. 
We then perform relativistic first principles calculations
for an Fe monolayer on W(110) and explicitly reveal the asymmetry of the spin-wave spectrum
in case of wave-vectors parallel to the (001) direction.
Furthermore, we quantitatively interpret our results in terms of 
a simplified spin-model by using calculated Dzyaloshinskii-Moriya vectors. 
Our theoretical prediction should inspire experiments to explore 
the asymmetry of spin-waves, with a particular emphasis on
the possibility to measure the Dzyaloshinskii-Moriya interactions in ultrathin films. 
\end{abstract}

\pacs{
71.15.Mb % Density functional theory, local density approximation
71.15.Rf % Relativistic effects
75.30.Ds % Spin waves
75.70.Ak % Magnetic properties of monolayers and thin films
}

\maketitle

It is by now well-established that relativistic effects play
a fundamental role in the magnetism of nanostructures, in 
particular, for thin films and finite deposited nanoparticles. 
Over the past two decades, a vast number of experimental 
and theoretical studies has been published to explore related 
phenomena such as magnetic anisotropies,   
spin-reorientation phase transitions,
and non-collinear magnetic orderings.\cite{PB_JPCM99,Skomski_JPCM03,JB_SSR06,VBL_RPP08,Wein_08}

The antisymmetric exchange interaction between two magnetic atoms, 
$E_{DM}={\bf D}_{ij} \left( {\bf M}_i \times {\bf M}_j \right)$, 
where ${\bf M}_i$ and ${\bf M}_j$ denote the spin-moments of the
atoms labeled by $i$ and $j$,
has been proposed 50 years ago by Dzyaloshinskii~\cite{Dzyalo_57-58}
and Moriya~\cite{Moriya_PR60}. 
The ${\bf D}_{ij}$ is called the Dzyaloshinskii-Moriya 
vector being identical to zero if the sites $i$ and $j$ experience
inversion symmetry. 
%Although the origin and the consequences of the 
%Dzyaloshinskii-Moriya interaction (DMI) have been continuously discussed 
%in context to spin-glasses,\cite{Fert-Levy,SGPS_JPC88} 
%magnetic semiconductors of pyrite structure~\cite{HO_PRB94,Ader_PRB01},
%or to weak ferromagnetism in cuprates \cite{SAE_PRB93}
%and in frustrated antiferromagnets,\cite{SK_PRL96,ECL_PRB02,CSZ_PTP02}
It has been put forward just about ten years ago that
an enhanced Dzyaloshinskii-Moriya interaction (DMI) at surfaces or interfaces can
give rise to novel phenomena in nanomagnetism such as 
to noncollinear interlayer coupling,\cite{Xia_PRB55,SB_PRL}
to unidirectional competing magnetic anisotropies,\cite{SOK_PRB98}
or to stabilization of non-collinear (chiral) magnetic orderings.\cite{CL_JMMM98,BR_PRL01}

A breakthrough on this field happened when the resolution of spin-polarized
scanning tunneling microscopy enabled to detect magnetic pattern
formation on the atomic scale in monolayer-thin films. 
Such periodic modulations 
have been observed for Mn monolayers deposited on W(110) and W(001) 
and, could successfully be interpreted in terms of a combination of
relativistic first principles calculations and a simple micromagnetic model 
as the consequence of large DM interactions.\cite{Bode_Nature07,Ferriani_PRL08}.
Using the same theoretical basis it was even possible to explain
the homochirality of the domain walls in two monolayers of Fe on W(110),
\cite{Heide_PRB08} in agreement with previous experimental
observation.\cite{Kubetzka_PRB03}

In this Letter, we investigate a consequence of the DM interactions on the 
spin-wave spectra in ultrathin films, not yet explored in the literature.
We argue that the chiral degeneracy of the spin-wave (SW) spectrum 
can be lifted due to the Dzyaloshinskii-Moriya interactions
and discuss under which conditions such a chiral asymmetry occurs.
Based on relativistic first principles calculations,
we explicitly evidence the asymmetry of the SW spectrum 
of an Fe monolayer on W(110)
in case of wave-vectors parallel to the (001) axis.
We then quantitatively interpret our results in terms of 
a simplified spin-model by using calculated DM vectors. 
%Since the chiral asymmetry is ultimately related to 
%a wave-vector asymmetry of the SW spectrum,
%$E({\bf q}) \ne E(-{\bf q})$,
By emphasizing the possibility of probing the DM interactions in 
ultrathin films, we impel experiments to explore the proposed effect. 

We start our study with simple considerations based on classical
spin-waves. If the atomic magnetic moment in the ground state 
of a ferromagnetic 
monolayer is ${\bf M}_0=M_0 {\bf e}_0$ with ${\bf e}_0$ being a unit vector,
then a spin-wave of wave-vector ${\bf q}$ and  chirality index 
(rotational sense) $c=\pm 1$
is defined by the magnetic orientations,
${\bf e}_i ({\bf q},c)$ = 
${\bf n}_1 \cos({\bf q} {\bf R}_i ) \sin\theta$
+ $c \,{\bf n}_2 \sin ({\bf q} {\bf R}_i) \sin\theta$ + 
${\bf e}_0 \cos\theta$, where 
${\bf n}_1 \perp {\bf e}_0$ and ${\bf n}_2 = {\bf n}_1 \times {\bf e}_0$ 
are unit vectors, 
${\bf R}_i$ is the position vector of site $i$ and 
the $\theta$ is the relative angle between the moments and ${\bf e}_0$.
Inspecting the energy of the SW in terms of an extended Heisenberg model containing
tensorial exchange interactions,\cite{Udvardi_PRB03} 
it turns out that only the antisymmetric exchange
interactions, i.e., the DM interactions give rise to a chirality dependent contribution,
\begin{equation}
E_{DM} ({\bf q},c) = c \sin^2\theta \sum_{i \ne j} \left( 
{\bf D}_{ij} \cdot {\bf e}_0 \right)
\sin\left( {\bf q} ( {\bf R}_i -  {\bf R}_j) \right) \; .
\label{eq:edm_sw}
\end{equation}
The above expression also implies that only the components of the DM vectors 
parallel to ${\bf e}_0$ influence the SW energy and that
a reversed chirality can be converted into a propagation of the SW
in the opposite direction: $E_{DM} ({\bf q}, -c) = 
E_{DM} (-{\bf q}, c) = - E_{DM} ({\bf q}, c)$.

The orientations of the DM vectors in a ferromagnetic monolayer have been analyzed 
for different 2D lattices in Refs.~\cite{CL_JMMM98} and \cite{Elena_PRB07}.
In particular, if the lattice has a twofold rotational symmetry,
%with respect to the axis at the bisecting point of the line between any pairs, 
such as in case of the (001) and (011) surfaces of cubic lattices, 
all the DM vectors lie in-plane. 
Clearly from Eq.~(\ref{eq:edm_sw}), a chiral asymmetry of the SW
occurs then only for an in-plane ground-state magnetization. Furthermore,
if the ground-state magnetization is in a mirror plane  of the monolayer 
no chiral asymmetry applies for wave-vectors being parallel with ${\bf e}_0$.

In order to demonstrate the chiral asymmetry of the SW's 
we have chosen a ferromagnetic Fe monolayer deposited on  W(110),
since $(i)$ it exhibits an in-plane ground-state 
magnetization~\cite{EHG_prb96,pratzer_prl01} and $(ii)$, 
as for the Fe double-layer~\cite{Heide_PRB08}
or for a Mn monolayer on W(110) and 
W(001),~\cite{Bode_Nature07,Udvardi_PhysicaB08,Ferriani_PRL08} 
large DM interactions are expected. 

\begin{figure}[ht]
\begin{center}
\includegraphics[width=4.5cm,bb=230 195 680 435,clip]{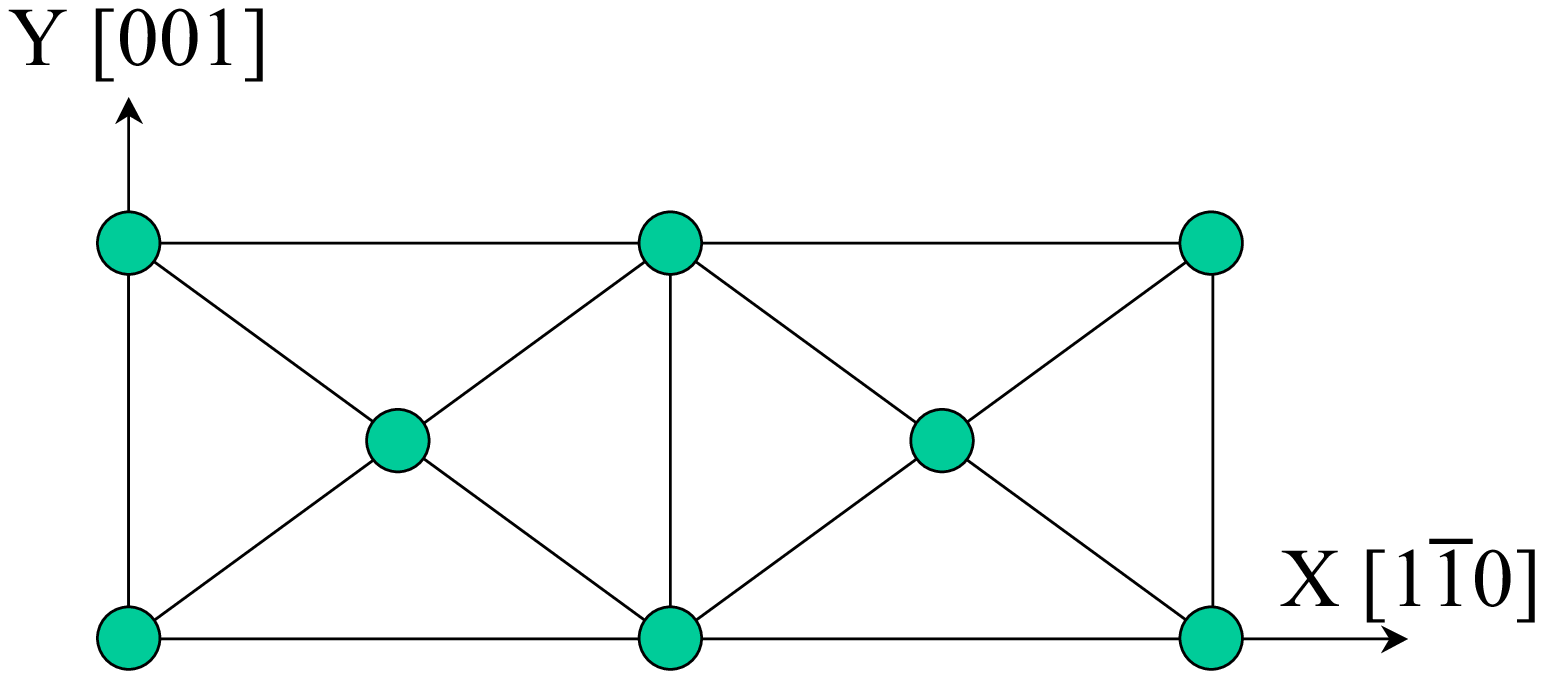}
\quad
\includegraphics[width=3.5cm,bb=295 180 580 425,clip]{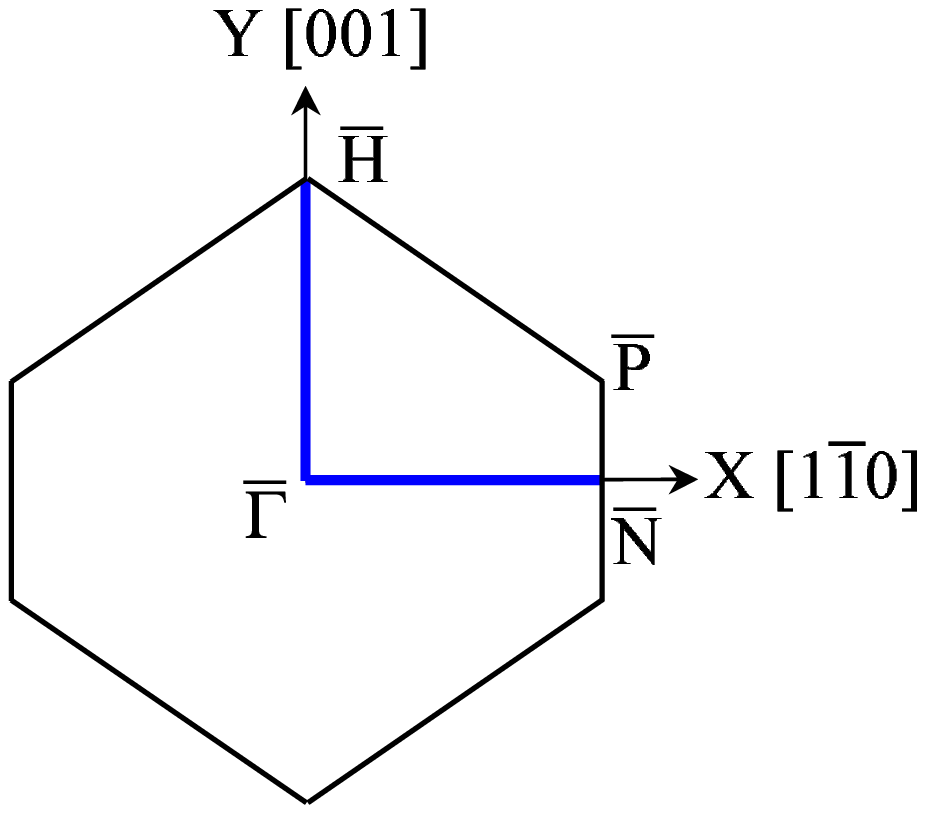}
\end{center}
\vskip -12pt 
\caption{(Color online) Sketch of the lattice positions (left) and of 
the surface Brillouin zone (right) of a bcc(110) plane. The high symmetry
points of the surface Brillouin zone are also labeled.
\label{fig:lattice}}
\end{figure}

Our notation used for the principle axes of a bcc(110) plane are shown 
in Fig.~\ref{fig:lattice} depicting the structure of the real-lattice 
and the surface Brillouin zone with the high symmetry points. 
The theoretical bcc W lattice constant as given in Ref.~\onlinecite{qh_prb99},
$a=3.205$ \AA, 
was chosen for the in-plane lattice constant (along the Y axis) throughout
the system. All the interlayer
distances were fixed to the ideal bcc(110) value, $d=\sqrt{2}a/2=2.266$ \AA, 
but the interlayer distance between the Fe and the topmost W layer 
was relaxed by -12.9 \%
($d_{Fe-W}=1.974$ \AA) according both to experiment~\cite{albrecht_ssc91}
and to theory~\cite{qh_prb03}. 
The calculations were performed in terms of the fully relativistic Screened 
Korringa-Kohn-Rostoker (SKKR) method~\cite{SKKR-book} by using the 
local density approximation and the atomic sphere approximation (ASA). 
It should be noted that the SKKR method makes use of a semi-infinite geometry
for the substrate, therefore, the calculations are
not affected by ambiguities related to a supercell or film geometry. 

We calculated magnetic
anisotropy energies, $E(001)-E(1\overline{1}0)= 2.11$ meV and
$E(110)-E(1\overline{1}0)= 0.41$ meV. This implies that, in agreement with
other theoretical works~\cite{ah_prb06,costa_prb08} and 
with the experiment,\cite{EHG_prb96}, the ground-state magnetization 
of FeW(110) is in-plane with an easy axis along the $(1\overline{1}0)$ direction
and the hard axis is along the $(001)$ direction. 
It should be noted that the magnetostatic dipole-dipole
interaction also favors the $(1\overline{1}0)$ direction by about 0.01 meV with respect 
to the $(001)$ axis and by 0.11 meV with respect to the $(110)$ direction.\cite{heide_diss}

We applied a recent relativistic extension~\cite{Udvardi_PRB03}
of the torque method~\cite{Lichtenstein_JMMM87} 
to evaluate tensorial exchange interactions for FeW(110) from first principles.
This method opened the way to atomistic spin-model simulations of 
nanostructures 
accounting for relevant relativistic interactions, such as
the on-site magnetic anisotropy, the anisotropic symmetric 
exchange interaction and the antisymmetric exchange 
interaction.\cite{Elena_PRB07,Udvardi_PhysicaB08,Antal_PRB08,Hubert_PRB08}

%Reassuringly, we obtained dominating ferromagnetic isotropic exchange interactions
%between the Fe atoms. 
By using the convention, $H=-\sum_{i\ne j} J_{ij} {\bf e}_i
{\bf e}_j$, our calculated isotropic exchange interactions
for the first few neighbors are,
$J_{01}=10.84$ meV, $J_{02}=-3.34$ meV, $J_{03}=3.64$ meV and $J_{04}=4.60$ meV.
Note that, in particular, the nearest neighbor interaction, 
$J_{01}$, is about four times less in
magnitude than the corresponding parameter in Ref.~\onlinecite{costa_prb08}.
To lend confidence to our values for $J_{ij}$, we performed Monte Carlo simulations
and obtained a Curie temperature of about 270 K, in very good agreement
with experiment (225 K).\cite{EHG_prb96} 
Note that random phase approximation
(RPA) calculations in Ref.~\onlinecite{costa_prb08} provided a $T_C$ 
above 1000 K which is,
most likely, the consequence of the overestimated NN exchange interaction. 

By using a canonical quantization of the linearized Landau-Lifshitz equations,
we also developed a method to calculate
the adiabatic SW spectra of bulk and layered systems  
on a relativistic first principles basis.\cite{Udvardi_PRB03} 
Although, within this approach, the interaction
of the spin-waves with the Stoner continuum is neglected, the main features 
of the SW spectra due to relativistic effects are expected to be well-described.
Notably, in case of a monolayer, two SW solutions are obtained with the energies,
$E^+({\bf q})$ and $E^-({\bf q})$, that correspond to the
chirality indices +1 and -1, respectively. 

\begin{figure}[ht]
\begin{center}
\includegraphics[width=7cm,bb=83 400 497 695,clip]{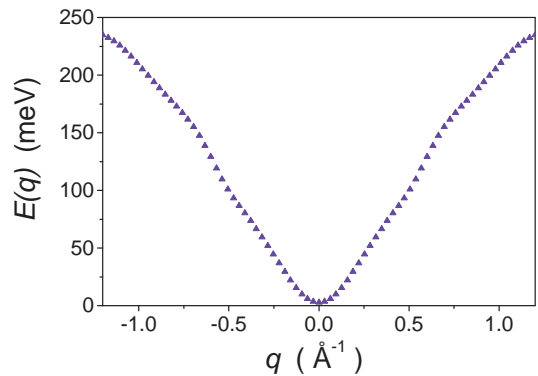}
\end{center}
\vskip -20pt 
\caption{(Color online) Calculated adiabatic spin-wave spectrum of FeW(110) along the X axis as given in Fig.~\ref{fig:lattice}.
\label{fig:spwx}}
\end{figure}

In Fig.~\ref{fig:spwx} the calculated adiabatic SW spectrum is shown along the X axis.
It was demonstrated in Ref.~\onlinecite{costa_prb08} 
that the adiabatic SW energies and 
the SW dispersion obtained from RPA agree well for wave-numbers as large as about 1 \AA$^{-1}$.
We, therefore, display the adiabatic SW spectrum for only $|q| < 1.2$  \AA$^{-1}$. Anticipated from
the symmetry analysis above, since in this case
the ground-state magnetization and the wave-vectors lie in a mirror plane
of the system, the spectrum is degenerate, i.e., $E^+(q)=E^-(q)$. 
Correspondingly, the SW dispersion is symmetric,
 $E(q)=E(-q)$. Note that the energy range of the dispersion in 
Fig.~\ref{fig:spwx} is approximately half of
that in  Ref.~\onlinecite{costa_prb08}, which we again attribute to
the very different exchange interaction parameters in the 
two theoretical works.

\begin{figure}[ht]
\begin{center}
\includegraphics[width=7cm,bb=92 400 500 690,clip]{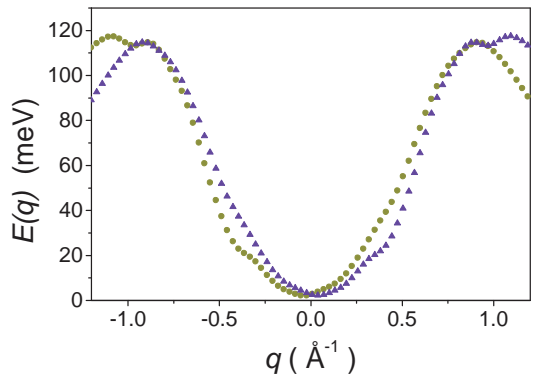}
\end{center}
\vskip -20pt 
\caption{(Color online) Calculated adiabatic spin-wave spectra with chirality
index +1 (triangles) and -1 (spheres) of FeW(110) along the Y axis, 
see Fig.~\ref{fig:lattice}.
\label{fig:spwy}}
\end{figure}

Next we inspect the SW spectrum for wave-vectors parallel to the (001) axis
displayed in Fig.~\ref{fig:spwy}. 
Since in this case ${\bf q}$ is perpendicular to the
ground-state magnetization, our symmetry analysis predicts lifting of the chiral degeneracy
of the spectrum, which can evidently be inferred from Fig.~\ref{fig:spwy}.
Furthermore, the relationship $E^+(-q)=E^-(q)$ is clearly regained.
As compared with Ref.~\onlinecite{costa_prb08}, again a difference by a factor
of two in the energy range of the magnons can be noticed. 

\begin{figure}[ht]
\begin{center}
\includegraphics[width=5cm,bb=180 80 1110 939,clip]{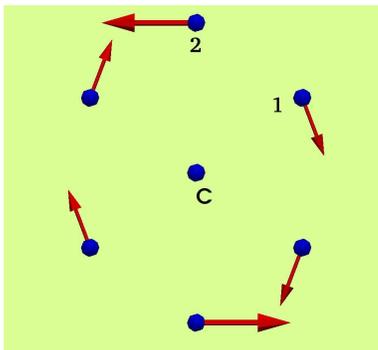}
\end{center}
\vskip -12pt 
\caption{(Color online) Sketch of the calculated Dzyaloshinskii-Moriya vectors 
between an atom (C)
and its nearest (1) and next nearest (2) neighbors in an Fe monolayer on W(110).
\label{fig:DMvecs}}
\end{figure}

In order to demonstrate that the observed asymmetry 
of the SW spectrum results
from the DM interactions we performed a model calculation for 
$\Delta E(q) = E^+(q)-E^-(q)$. 
Our first principles calculations indicated
that the DM vectors for the nearest and next nearest neighbors,
visualized in Fig.~\ref{fig:DMvecs},
are at least by an order larger in magnitude
than the ones for more distant pairs. 
%It should, however, be mentioned that the DM interactions decay quite slowly
%with the distance.\cite{Hubert_PRB08}
By using the method described in Ref.~\onlinecite{Udvardi_PRB03} 
the asymmetry of the SW energy can then be expressed as,
\begin{equation}
\Delta E(q) = \frac{16 \mu_B}{M_0} D^x_1 \sin(\frac{1}{2}qa)-\frac{8 \mu_B}{M_0} D^x_2 \sin(qa)
\; ,
\label{eq:deq}
\end{equation}
where $M_0=2.22 \, \mu_B$ is the spin-magnetic moment per atom and  $D^x_1$ and 
$D^x_2$ are the magnitudes of the $x$ components 
(parallel to the ground-state magnetization) 
of the DM vectors for the nearest and second nearest neighbors, respectively.

\begin{figure}[ht]
\begin{center}
\includegraphics[width=7cm,bb=85 395 500 695,clip]{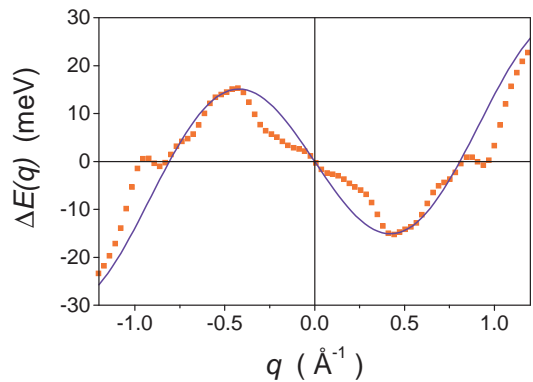}
\end{center}
\vskip -20pt 
\caption{(Color online) Squares: asymmetry of the spin-wave spectrum of
FeW(110) along the Y axis as derived from the data in Fig.~\ref{fig:spwy}, 
solid line: the function, Eq.~(\ref{eq:deq}), obtained from a second nearest neighbor
model with the calculated DM interactions.
\label{fig:spwyasym}}
\end{figure}

In Fig.~\ref{fig:spwyasym} we plotted the asymmetry, $\Delta E(q)$, of the SW spectrum 
of FeW(110) along the Y axis obtained from the data in Fig.~\ref{fig:spwy}.
As can be inferred from this figure, $\Delta E(q)$ exhibits
local extrema at about $q = \pm$0.44 \AA$^{-1}$ with $|\Delta E(q)| \simeq $ 15 meV and
changes sign at $q = \pm$0.83 \AA$^{-1}$.
Apparently, these features of $\Delta E(q)$ are fairly well reproduced by the function,
 Eq.~(\ref{eq:deq}), when using the calculated parameters,  $D^x_1=1.42$ meV and $D^x_2=6.08$ meV.
Note that the characteristic extrema of $\Delta E(q)$ are determined by the DM
interactions between the next nearest neighbors, since the $\sin(qa)$ function in the
second term on the {\em rhs} of  Eq.~(\ref{eq:deq}) reaches a maximum/minimum at $|q|=\pi/2a \simeq
0.49$ \AA$^{-1}$.
The deviations of the asymmetry of the SW energy from this model function  are related
to the DM interactions between more distant pairs that add low-frequency
modulations to the SW dispersion.

The magnon spectrum of FeW(110) along the Y direction has been measured very recently 
by using spin-polarized electron energy loss spectroscopy (SPEELS),\cite{prokop_08}
a highly suitable technique to probe high wave-vector magnetic excitations of ultrathin films.
Surprisingly, the measured magnon energies are about half of the theoretical values 
reported here and smaller by even a factor of four than the
calculated values in Ref.~\onlinecite{costa_prb08}.
We are, however, aware of linear response calculations~\cite{buczek} that
provided with a very similar magnon dispersion along the Y axis
as compared to that in Fig.~\ref{fig:spwy}.
Thus, the low energy of the measured magnon spectrum~\cite{prokop_08} 
should most probably be attributed to effects not 
included in the first principles calculations, 
such as spin-charge coupling~\cite{ps_prb08}
or phonon-magnon interaction.\cite{lazewski_prb07}

Considering the size of the SW asymmetry obtained from our calculations, 
e.g., about 20~\% at
$q = \pm$0.44 \AA$^{-1}$ with respect to the average energy, $(E^+(q)+E^-(q))/2$,
we strongly suggest that it should be accessible to experiments. 
Indeed, preliminary measurements on FeW(110)~\cite{prokop}
indicate the presence of an asymmetry in the magnon spectrum being quite similar 
in size and shape as in Fig.~\ref{fig:spwyasym}. 
%Several sources for the occurrence of such an asymmetry, other than proposed in the present Letter, must, however, be ruled out.
Further candidates for experimental observation of the proposed SW asymmetry
are ferromagnetic monolayers on substrates
with large spin-orbit coupling and polarizability (W, Pt or Ir). 
In case of an out-of-plane ground-state magnetization,
a small magnetic field might be applied to orient the magnetization in-plane, in order to fulfill
the necessary condition for the chiral asymmetry of magnons.
Our model calculation, see Eq.~(\ref{eq:deq}), clearly implies that
such experiments would serve as a unique tool to measure
the Dzyaloshinskii-Moriya interactions in ultrathin ferromagnetic films,
to be directly compared with the results of ab initio calculations. 
Concerning, in particular, the role of relativistic effects,
such a progress would clearly assist a deeper understanding of the magnetism in
nanostructures.
 
The authors appreciate useful discussions with J. Prokop and J. Kirschner. 
Financial support of the Hungarian National Scientific
Research Foundation (OTKA contracts No. OTKA K68312, No. K77771 and No. NF61726)
is acknowledged.

\bibliographystyle{myst}

\end{document}